\documentclass[preprint,showpacs,preprintnumbers,amsmath,amssymb]{revtex4}

\usepackage{graphicx}
\usepackage{dcolumn}
\usepackage{bm}

\begin{document}

\preprint{APS/123-QED}

\title{Coherence Transition of Small Josephson Junctions Coupled to a Single-Mode
Resonant Cavity: Connection to the Dicke Model}

\author{Kohjiro Kobayashi and David Stroud}
\affiliation{%
Department of Physics, Ohio State University, Columbus, OH 43210\\
}%

\date{\today}

\begin{abstract}

We calculate the thermodynamic properties of a collection of $N$ small Josephson junctions coupled to a single-mode resonant electromagnetic cavity, at finite temperature $T$, using several approaches. In the first approach, we include all the quantum-mechanical levels of the junction, but treat the junction-cavity interaction using a mean-field approximation developed previously for $T = 0$. In the other approaches, the junctions are treated including only the two lowest energy levels per junction, but with two different Hamiltonians. The first of these maps onto the Dicke model of quantum optics. The second is a modified Dicke model which contains an additional XY-like coupling between the junctions. The modified Dicke model can be treated using a mean-field theory, which in the limit of zero XY coupling gives the solution of the Dicke model in the thermodynamic limit using Glauber coherent states to represent the cavity. In all cases, for an $N$-independent junction-cavity coupling, there is a critical junction number $N$ above which there is a continuous transition from incoherence to coherence with decreasing $T$. If the coupling scales with $N$ so as to give a well-behaved thermodynamic limit, there is a critical minimum coupling strength for the onset of coherence. In all three models, the cavity photon occupation numbers have a non-Bose distribution when the system is coherent.


\end{abstract}

\maketitle

\section{Introduction}
When a two-dimensional array of Josephson junctions is driven by an applied current, it can radiate coherently. Experiments showing this behavior have emphasized current-driven arrays of overdamped Josephson junctions \cite{Jain1984, Benz1991}. The radiated coherent power from such arrays has been predicted to be proportional to the square of the number $N$ of Josephson junctions in the array \cite{Tilley1970}. More recently, coherent emission from {\em underdamped} one-dimensional Josephson arrays coupled to a single-mode electromagnetic cavity has been experimentally studied \cite{Barbara1999, Cawthorne1999, Vasilic2001, Vasilic2002}. In this work, it was shown that no coherent radiation is emitted below a threshold number $N_c$ of junctions, but above this threshold the array can radiate coherently, with emitted power again proportional to $N^2$. Such behavior had already been predicted much earlier, on the basis of an analogy between a one-dimensional voltage biased series array and a collection of two-level atoms coupled to an electromagnetic cavity \cite{Bonifacio}; the analogy suggests that this radiation is the Josephson analog of the population inversion that leads to coherent emission in a laser.

A simple model Hamiltonian to describe this coherent radiation, taking into account the quantum-mechanical nature of both the junctions and the cavity, was suggested recently \cite{Harbaugh2000}. In this paper, the ground state of the model Hamiltonian is obtained within a mean-field theory (MFT). In agreement with experiment, the MFT predicts that there is a critical threshold number $N_c$ of junctions for the onset of coherence at fixed coupling strength.

The mechanism for coherent radiation from a Josephson junction array resembles that of superradiance in a system of N two-level atoms coupled to a electromagnetic field \cite{Eastham2003}. The latter system can be treated the Dicke model \cite{Dicke1954}, which describes the system of identical two-level atoms in a single-mode radiation field. Emission and absorption within the Dicke model have been extensively studied \cite{Cummings1983, Cummings1986}. The predicted response of this two-level atom/radiation system agrees qualitatively with that of an array of Josephson junctions \cite{Alsaidi2001, Alsaidi2002}. It has also been shown \cite{Alsaidi2002}, that a modified Dicke Hamiltonian, which contains an additional term resembling a dipole-dipole interaction between the junctions, is a better approximation to the cavity-junction system than is the original Dicke model.

In contrast to the Josephson/cavity system, the Dicke model can be solved in the thermodynamic limit, i.e. $N\rightarrow\infty$, volume $V$ $\rightarrow\infty$, and $N/V \rightarrow$ const \cite{Hepp1973,Wang1973,Hioe1973}. The solution yields a continuous transition from a normal to a superradiant state at a critical coupling strength, for fixed temperature $T = 0$.

In this paper, we extend the model \cite{Harbaugh2000} to finite $T$. We find, again within MFT, that there is a critical threshold number $N_c(T)$ for coherence at a sufficiently low $T$. If $T$ is increased at fixed $N > N_c(0)$, there is a continuous transition from coherence to incoherence at an $N$-dependent temperature $T_c(N)$. To test the MFT, we compare its predictions with those of the Dicke model \cite{Wang1973, Hioe1973} and of the modified Dicke
model \cite{Alsaidi2002}. Our Hamiltonian does not map exactly onto these models, because the individual Josephson junctions have more than two quantum levels, whereas the Dicke and modified Dicke models assume two-level systems interacting with a single-mode cavity. Nonetheless, when the parameters of our model systems are such that the lowest two levels of the junction are well-separated from the higher levels, the MFT agrees well with the two-level model predictions. We also show that, when applied to the Dicke model, the MFT is equivalent to a coherent state expansion, an approach known to give the solution of the Dicke model at large $N$.

The remainder of this paper is organized as follows. In Section \ref{sec:jose_model}, we describe the MFT for N Josephson junctions interacting with a cavity at finite $T$. In Section \ref{sec:jose_dicke}, we review the coherent state treatment of the Dicke model, present a MFT for both the Dicke model and the modified Dicke model, and show that, when applied to the Dicke model, the MFT is equivalent to the coherent state approach. In Section \ref{sec:jose_result}, we give numerical results for all three models. In Section \ref{sec:jose_limit}, we show how the Josephson-cavity model of Section \ref{sec:jose_model} can be mapped onto the Dicke model when the Josephson coupling is small compared to the charging energy; we also discuss how the parameters of all three models must scale in the thermodynamic limit. Section \ref{sec:jose_discussion} presents a concluding discussion.

\section{Many Josephson Junctions Interacting with a Single-mode Cavity \label{sec:jose_model}}
\subsection{Model Hamiltonian}

An array of $N$ underdamped voltage-biased Josephson junctions in a lossless electromagnetic cavity having a single electromagnetic mode of frequency $\omega$ may be described by the following idealized model Hamiltonian:
\begin{equation}
H=H_{photon}+\sum_{j=1}^N H_{Jj}. \label{eq:H}
\end{equation}
Here \begin{equation} H_{photon} = \hbar\omega
(a^{\dag}a+\frac{1}{2})
\end{equation}
is the photon Hamiltonian, $a^{\dag}$ and $a$ being the photon creation and annihilation operators for photons having angular frequency $\omega$, which satisfy the usual commutation relations. $[a,a^{\dag}]=1,[a,a]=[a^{\dag},a^{\dag}]=0$. The Hamiltonian of the
j$^{th}$ Josephson junction can be expressed as
\begin{equation}
H_{Jj} = \frac{1}{2}U(n_j-\bar{n}_j)^2-J \cos{\gamma_j}.
\label{eq:hj}
\end{equation}
Here $U=4e^2/C$ is the capacitive energy of the junction, $e$ is the electronic charge, $C$ is the junction capacitance. $n_j$ is an operator representing the difference in the number of Cooper pairs on the two superconducting islands forming the junction, $\bar{n}_j$ is related to the gate voltage across the j$^{th}$ junction, $J=\hbar I_c/(2e)$ is the Josephson coupling energy of the junction, $I_c$ is the junction critical current, and finally $\gamma_j$ is
the gauge-invariant phase difference across the junction.

Explicitly, $\gamma_j$ may be written
\begin{equation}
\gamma_j=\phi_j-\frac{2\pi}{\Phi_0}\int_j{\bf A}\cdot d{\bf l},
\label{eq:gamma}
\end{equation}
where $\phi_j$ is the phase difference across the junction in a
particular gauge, and ${\bf A}$ is the vector potential due to the
cavity mode (given explicitly below) in the same gauge,
$\Phi_0=hc/(2e)$ is the flux quantum, and the line integration is
carried out across the junction.  The operators $n_j$ and $\phi_j$ are canonically conjugate and satisfy the commutation relation $[n_k,\phi_l]=i\delta_{kl}$, which is satisfied if we use the representation $n_k =i\frac{\partial}{\partial\phi_k}$.

In the Coulomb gauge, $\nabla\cdot{\bf A}=0$, ${\bf A}$ can be expressed as
\begin{equation}
{\bf A}=\sqrt{\frac{hc^2}{\omega V}}(a+a^{\dag}){\bf E}({\bf x}),
\label{eq:vectorp}
\end{equation}
where $V$ is the cavity volume, and ${\bf E}({\bf x})$ is proportional to the local electric field of the cavity mode, normalized so that $\int_V|{\bf E}({\bf x})|^2d^3x = 1$. Introducing a coupling parameter
\begin{equation}
g_j=\frac{2\pi}{\Phi_0}\sqrt{\frac{hc^2}{\omega V}}\int_j{\bf
E}({\bf x})\cdot d{\bf l}, \label{eq:gj}
\end{equation}
we may rewrite $\gamma_j$ as
\begin{equation}
\gamma_j = \phi_j-g_j(a+a^{\dag}).
\end{equation}
Eq.\ (\ref{eq:gj}) suggests that typically $g_j \propto 1/\sqrt{V}$ for a given mode (provided that the cavity {\em shape} does not change as the volume increases.

\subsection{\label{sec:MFT} Mean-field approximation}
We now develop a suitable mean-field approximation for the Hamiltonian (\ref{eq:H}), for both zero and finite $T$. We consider only the case of identical Josephson junctions, so that all $g_j=g$ and $\bar{n}_j = \bar{n}$; the extension to non-identical junctions is straightforward \cite{Kobayashi}. We also assume that the coupling parameters $g$ are weak. In this case, we can expand the cosine in eq.\ (\ref{eq:hj}), retaining only the term of first order in $g(a+a^{\dag})$,
so that $\cos{\gamma_j} \sim \cos{\phi_j}+g(a+a^{\dag})\sin\phi_j$.
Within this approximation, the only part of the Hamiltonian that depends on both cavity and junction variables is
\begin{equation}
H_{int}\sim -gJ(a+a^\dag)\sum_j\sin\phi_j.
\label{eq:Hint}
\end{equation}

The eigenvalues and eigenfunctions of $H$ can now be found if we make the following mean-field approximation for $H_{int}$:
\begin{eqnarray}
H_{int} \sim H_{int}^m \equiv
&-& gJ(a+a^\dag)\sum_j\langle\sin\phi_j\rangle \nonumber \\
&-& gJ\langle a+a^\dag\rangle\sum_j\sin\phi_j \nonumber \\
&+& gJ\langle a+a^\dag \rangle\sum_j\langle\sin\phi_j\rangle.
\label{eq:Hintm}
\end{eqnarray}
Here $\langle ... \rangle$ denotes a canonical average at temperature $T$ with respect to the mean-field Hamiltonian $H^m$.
$H^m$ is now given by
\begin{equation}
H^m=H_{photon}^m+\sum_j H_{Jj}^m+H_{c}^m,
\end{equation}
where
\begin{equation}
H_{photon}^m  = \hbar\omega(a^{\dag}a+\frac{1}{2})-g J(a+a^{\dag})
\sum_j\langle\sin{\phi_j}\rangle ,\label{eq:Hmphoton}
\end{equation}
\begin{equation}
 H_{Jj}^m  =
\frac{U}{2}(n_j-\bar{n})^2-J\cos{\phi_j}-gJ\langle a+a^{\dag}\rangle
\sin{\phi_j} ,\label{eq:HmJ}
\end{equation}
and
\begin{equation}
 H_{c}^m = g J \langle
a+a^{\dag}\rangle\sum_j\langle\sin{\phi_j}\rangle. \label{eq:Hmc}
\end{equation}
Evidently, the first term depends only on the photon variables, the second is a sum of single-junction terms, and the third is simply a c-number.

We now introduce the variable $\lambda_j=\langle\sin{\phi_j}\rangle
= \lambda$, since $\lambda_j$ is independent of $j$. In terms of
$\lambda$, we may rewrite the photon term (\ref{eq:Hmphoton}) as
\begin{equation}
H_{photon}^m =
\hbar\omega(a^{\dag}a+\frac{1}{2})-gJ(a+a^{\dag})N\lambda.
\label{eq:hphotmf}
\end{equation}
This is simply the Hamiltonian of a {\em displaced} harmonic oscillator. Its eigenvalues $E_n$ and normalized eigenfunctions $\psi_n(x)$ are just $E_n =
\hbar\omega(n+\frac{1}{2})-\frac{J^2g^2N^2\lambda^2}{\hbar\omega}$
and $\psi_n(x) =  \langle x|e^{-ip\langle
x\rangle/\hbar}|n\rangle=\frac{1}{\sqrt{2^nn!}}\left(\frac{\omega}{\pi\hbar}\right)^{1/4}
e^{-\frac{\omega}{2\hbar}(x-\langle x\rangle)^2}H_n
\left(\sqrt{\frac{\omega}{\hbar}}(x-\langle x\rangle)\right)$. Here
$H_n$ is a Hermite polynomial,
$p=i\sqrt{\frac{\hbar\omega}{2}}(a^{\dag}-a)$
is a momentum operator, $\langle x\rangle =
\frac{gJN\lambda}{\omega}\sqrt{\frac{2}{\hbar\omega}}$ is the mean
displacement, and we have used the relation $\langle
a+a^{\dag}\rangle=\frac{2JgN\lambda}{\hbar\omega}$. The canonical
partition function corresponding to $H_{photon}^m$ is just
$Z_{photon}^m=\frac{e^{\beta
\xi^2/(\hbar\omega)}}{2\sinh{\beta\hbar\omega/2}}$, where
$\xi=gJN\lambda$ and $\beta = 1/(k_BT)$.

Next, we consider the Josephson junction Hamiltonian $H_{Jj}^m$
[eq.\ (\ref{eq:HmJ})]. With the definitions $K(\lambda)=J\sqrt{1+4\left(\frac{g^2JN\lambda}{\hbar\omega}\right)^2}$
and $\psi(\phi_j)=e^{-i\bar{n}(\phi_j-\alpha)}u(\phi-\alpha)$, where
$\tan{\alpha}=\frac{2Jg^2N\lambda}{\hbar\omega}$, the
Schr\"{o}dinger equation for the junctions, $H_{Jj}\psi(\phi_j) =
E\psi(\phi_j)$, reduces to the standard Mathieu equation
\cite{Mathieu}:
\begin{equation}
\frac{d^2y(v_j)}{dv_j^2}+(a-2q\cos{2v_j})y(v_j)=0,
\label{eq:Mathieu}
\end{equation}
where $v_j=(\phi_j-\alpha)/2$, $y(v_j)=u((\phi_j-\alpha)/2)$,
$q=-4K(\lambda)/U$, the characteristic value of the Mathieu equation is $a=8E/U$, and we have used the representation $n_j=i\frac{\partial}{\partial\phi_j}$. The allowed eigenvalues are determined by the condition that $\psi(\phi_j+2\pi) = \psi(\phi_j)$, or equivalently, that $y(v_j+\pi)=\exp(2i\bar{n}\pi)y(v_j)$. The allowed solutions $y_{\nu}(v_j)$ are therefore the Floquet (Bloch) functions of $v_j$,
with Floquet exponent $\nu=2\bar{n}+2k$, where $k=0,\pm1,\pm2,...$.
The corresponding eigenvalues of $H_{Jj}$ are labeled by the quantum
number $\nu = 2\bar{n}+2k$ and the parameter $q$, and may be denoted $E(\nu=2\bar{n}+2k;q)$. For $0\le\bar{n}\le 0.5$, the lowest eigenvalue corresponds to $k=0$, followed in order by $k=-1,1,-2,2,..$. Including only these Floquet solutions, we can formally express the junction partition function as
\begin{equation}
Z_{Jj}^m=\sum_{k=0,\pm1,\pm2,...}e^{-\beta E(2\bar{n}+2k;q)} \equiv
Z_J^m,
\end{equation}
where the last identity holds for identical junctions.

We now determine the properties of the junction-cavity system within MFT. At $T = 0$, the system is in its ground state, and the approximate ground state properties can be obtained analytically as shown in Ref.\ \cite{Harbaugh2000}. For $-0.5\le\bar{n}\le 0.5$, the ground state energy is
\begin{equation}
E_g(\lambda)=\frac{\hbar\omega}{2}+\frac{(gJN\lambda)^2}{\hbar\omega}+N\frac{U}{8} a(\nu=2\bar{n};q) ,
\label{eq:e0}
\end{equation}
where $a(\nu;q)$ is the eigenvalue of eq.\ (\ref{eq:Mathieu}) corresponding to characteristic exponent $\nu$ and parameter $q$. If we use the approximate analytical expression \cite{Harbaugh2000}
\begin{equation}
a(\nu=2\bar{n};q)\sim
4\left(1-4\bar{n}^2-\sqrt{(1-4\bar{n}^2)^2+\frac{q^2}{4}}\right)+4\bar{n}^2,
\label{eq:analyt}
\end{equation}
we obtain
\begin{equation}
E_g(\lambda)\sim\frac{\hbar\omega}{2}+\frac{(gJN\lambda)^2}
{\hbar\omega}+N\frac{\bar{U}}{2}\left\{1-
\sqrt{1+\frac{4J^2}{\bar{U}^2}\left[1+4\left(\frac{g^2JN\lambda}{\hbar\omega}\right)^2\right]}\right\}+N
\frac{U\bar{n}^2}{2} ,
\end{equation}
where $\bar{U}=U(1-4\bar{n}^2)$.  $\lambda$ is determined by the condition $\frac{d E_0(\lambda)}{d\lambda}=0$, which leads to
\begin{equation}
\lambda^2(T = 0)\sim
1-\left(\frac{\hbar\omega}{2g^2JN}\right)^2\left(1+\frac{\bar{U}^2}{4J^2}\right).
\label{eq:lammin}
\end{equation}
Because the right hand side of this equation must be non-negative, the critical junction number $N_c(0)$ for a non-zero $\lambda$ at $T=0$ is
\begin{equation}
N_c(T= 0) \sim\frac{\hbar\omega}{4g^2J^2}\sqrt{\bar{U}^2+ 4 J^2}.
\label{eq:nc0}
\end{equation}
When $N\le N_c(0)$, $\lambda = 0$ corresponds to a minimum of the energy, but when $N\ge N_c(0)$, $\lambda = 0$ is a local maximum; the energy minimum occurs at $\lambda \neq 0$.

More generally, the exact solution for $N_c(T = 0)$ can be calculated from eq.\ (\ref{eq:e0}), supplemented by the condition $\frac{dE_g(\lambda)}{d\lambda} = 0$. The result is
\begin{equation}
\lambda (T = 0)=\frac{\sqrt{(a'g^2JN)^2-(\hbar\omega)^2}}{2g^2JN} ,
\end{equation}
where $a'(\lambda)=\frac{da(\nu=2\bar{n};q)}{dq}|_{q=-\frac{4K(\lambda)}{U}}$.
The corresponding critical number is
\begin{equation}
N_c(T = 0)=\frac{\hbar\omega}{|a'(0)|g^2J}, \label{eq:nc0e}
\end{equation}
where $a'(0)=a'(\lambda)|_{\lambda=0}$.

For $T\neq 0$, because $H^m$ is the sum of several commuting terms, the total partition function  $Z^m=Z_{photon}^m(Z_{J}^m)^NZ_{c}^m$, where $Z_{c} =\exp(-\beta H_{c}^m)$. The corresponding Helmholtz free energy $F^m$ is
\begin{equation}
F^m=-kT\ln{Z^m}= F_{photon}^m + NF_J^m + F_{c}^m, \label{eq:Fm}
\end{equation}
where $F_{photon}^m=kT\ln{\left(2\sinh{(\beta\hbar\omega/2)}\right)}-
\frac{(gJN\lambda)^2}{\hbar\omega}$, $F_J^m = -k_BT\ln Z_J^m$, and
$F_{c}^m = 2\frac{(gJN\lambda)^2}{\hbar\omega}$. When the coherence order parameter $\lambda \neq 0$, the Helmholtz free energy, for fixed $g$, is quadratic in the number of the junctions $N$. This quadratic dependence is a hallmark of the coherent state.

The actual value of $\lambda$ is obtained from the Helmholtz free energy, using the condition
\begin{equation}
\frac{dF^m(\lambda)}{d\lambda}=0. \label{eq:dF0}
\end{equation}
We have obtained $\lambda$ by solving eqs.\ (\ref{eq:Fm}) and (\ref{eq:dF0}) self-consistently. These equations may allow for several possible values of $\lambda$, of which we choose that value which gives the lowest $F^m$.

In all the above discussion, we have assumed implicitly that $g$ is independent of N. The expected behavior when $g$ depends on N is discussed below.

\section{Dicke Model and Generalized Dicke Model \label{sec:jose_dicke}}
\subsection{Model Hamiltonians}
In the previous section, we described a simple mean-field approximation for the statistical mechanics of the junction-cavity system. This approximation includes all the junction levels, but treats the junction-cavity interaction only approximately. We now describe an alternative approach, which retains only the two lowest energy levels of each junction. In this case, the Hamiltonian reduces to the well-known Dicke model of quantum optics. The Dicke Hamiltonian \cite{Dicke1954} is a simple model describing the interaction of N two-level systems with a single harmonic oscillator mode. It can be written (omitting the cavity zero-point energy)
\begin{equation} H^{Dicke}=\hbar\omega
a^{\dag}a+\sum_j\left(\frac{1}{2}\epsilon_j\sigma_z^j+\xi_j
a^{\dagger}\sigma_-^j+\xi_j^*a\sigma_+^j\right). \label{eq:Dicke}
\end{equation}
Here $\epsilon_j > 0$ is the energy level splitting of the $j^{th}$ two-level system at the j$^{th}$ junction, and $\xi_j$ is a parameter characterizing the strength of the coupling between the harmonic oscillator and the j$^{th}$ two-level system. The quantities $a^\dag$ and $a$ are raising and lowering operators, as above. The quantities $\sigma_\alpha^j$ are Pauli spin-1/2 spin operators and satisfy $[\sigma_\alpha^{j},\sigma_\beta^{k}]=2i\delta_{jk}\sigma_\gamma^i$, where $\alpha$, $\beta$, and $\gamma$ are cyclic commutations of (x,y,z). In order for the thermodynamic limit to exist, we must assume that $\xi_j \propto 1/\sqrt{N}$ for large $N$.

Besides the Dicke model, we also consider a modified Dicke model\cite{Alsaidi2002}, which is an extension of effective two-qubit model\cite{Joshi1991, Zheng2000, Blais2004} to the case of N coupled two-level systems,
\begin{equation} H^{M Dicke}=\hbar\omega
a^{\dag}a+\sum_j\left(\frac{1}{2}\epsilon_j\sigma_z^j+\xi_j
a^{\dagger}\sigma_-^j+\xi_j^*a\sigma_+^j\right)+\sum_{\langle
jk\rangle}\Omega_{jk}(\sigma_+^j\sigma_-^k+\sigma_-^j\sigma_+^k),
\label{eq:MDicke}
\end{equation}
where the last sum runs over all distinct pairs $jk$ (i.\ e., not including $j = k$). The last term in eq.\ (\ref{eq:MDicke}) is an effective direct junction-junction interaction. As discussed in Ref.\ \cite{Alsaidi2002}, the Hamiltonian (\ref{eq:MDicke}) generally gives levels in closer agreement with the Hamiltonian (\ref{eq:H}) than does the pure Dicke Hamiltonian (\ref{eq:Dicke}), when there is more than one photon excited in the cavity. A simple derivation of this term is given in Ref.\ \cite{Alsaidi2002}.

In order for the thermodynamic limit to exist in the modified Dicke model, we require not only that $\xi_j \propto 1/\sqrt{N}$, and but also that $\Omega_{jk} \propto 1/(N-1)$, as further explained below.

\subsection{\label{sec:Dicke}Statistical mechanics of Dicke model using Glauber coherent state expansion}

The thermodynamics of the Hamiltonian (\ref{eq:Dicke}) can be calculated in the limit $N \rightarrow \infty$, using a product basis consisting of the Glauber coherent states $|\alpha\rangle$ for the photons and eigenstates of $\sigma_z^i$ for the two-level systems \cite{Wang1973, Hioe1973}. In this section, we briefly review this solution.

The states $|\alpha\rangle$ are eigenstates of the lowering operator, i.\ e.,
\begin{equation}
a|\alpha\rangle = \alpha|\alpha\rangle.
\end{equation}
where the eigenvalue $\alpha$ is generally complex, since $a$ is a non-Hermitian operator. The eigenfunctions satisfy the completeness relation
$(1/\pi)\int_{-\infty}^\infty\int_{-\infty}^\infty
dRe(\alpha)dIm(\alpha)|\alpha\rangle\langle\alpha |=1$, the integral running over the entire complex $\alpha$ plane.

In terms of this basis, the partition function of the Dicke Hamiltonian takes the form
\begin{eqnarray}
Z_{Dicke} &=& \mathrm{Tr}e^{-\beta H^{Dicke}} \\
& =& \sum_{\sigma_1=\pm1}...\sum_{\sigma_N=\pm1}\int_{-\infty}^\infty\int_{-\infty}^\infty
\frac{dRe(\alpha)dIm(\alpha)}{\pi}
\langle\sigma_1...\sigma_N|\langle\alpha |
e^{-\beta H^{Dicke}}|\alpha\rangle |\sigma_1...\sigma_N\rangle \nonumber \\
 &=& \int_{-\infty}^\infty\int_{-\infty}^\infty\frac{dRe(\alpha)dIm(\alpha)}{\pi}
 e^{-\beta\hbar\omega | \alpha|^2}
 \prod_j\left(\sum_{\sigma_j =\pm 1} \langle\sigma_j |e^{-\beta
h_j}|\sigma_j\rangle\right) \label{eq:zdicke1}
\end{eqnarray}
where
\begin{equation}
h_j = \frac{1}{2}\epsilon\sigma^j_z+\xi_j
\alpha^*\sigma^j_-+\xi_j^*\alpha\sigma^j_+.
\end{equation}

In order to evaluate the sums, following Refs.\ \cite{Wang1973} and \cite{Hioe1973}, we expand the operator $\exp(-\beta H^{Dicke})$ in a Taylor series, assume that $a/\sqrt{N}$ and $a^{\dag}/\sqrt{N}$ exist in the limit N $\rightarrow\infty$, and finally, within individual terms of the Taylor expansion, interchange the order of the double limits as follows:
\begin{equation}
 \lim_{N\rightarrow\infty}\lim_{R\rightarrow\infty}\sum_{r=0}^R\frac{(-\beta H^{Dicke})^r}{r!}=
 \lim_{R\rightarrow\infty}\sum_{r=0}^R\lim_{N\rightarrow\infty}\frac{(-\beta H^{Dicke})^r}{r!} .
\label{eq:interchange}
\end{equation}

Each sum in eq.\ (\ref{eq:zdicke1}) can easily be evaluated, since it is just the trace of the operator $\exp(-\beta h_j)$, and the resulting partition function can be expressed as
\begin{eqnarray}
Z_{Dicke} &=& \int\frac{dRe(\alpha)dIm(\alpha)}{\pi}
e^{-\beta\hbar\omega|\alpha|^2}\prod_j{2\cosh{\left(\frac{\beta\epsilon_j}{2}
\sqrt{1+\left(\frac{4|\xi_j||\alpha |}{\epsilon_j}\right)^2}\right)}} \nonumber \\
 &=& 2\int_0^{\infty}rdr e^{-\beta\hbar\omega r^2}
 \prod_j\left(2\cosh{\frac{\beta\epsilon_j}{2}
 \sqrt{1+\left(\frac{4|\xi_j|r}{\epsilon_j}\right)^2}}\right) ,
\end{eqnarray}
where we have introduced $r=|\alpha|$ and written
$\int_{-\infty}^\infty\int_{-\infty}^\infty
\frac{dRe(\alpha)dIm(\alpha)}{\pi}=2\int_0^{\infty}rdr$.

To complete the evaluation of the free energy, we make the changes of variables $|\xi_j|=\frac{\xi_j^\prime}{\sqrt{N}}$ and $y=\frac{r^2}{N}$. Then $Z_{Dicke}$ takes the form
\begin{equation}
Z_{Dicke}=N\int_0^{\infty}dy \exp\left[N\phi(y)\right] ,
\end{equation}
where
\begin{equation}
\phi(y)=\left(-\beta\hbar\omega
y+\frac{1}{N}\sum_j\ln{(2\cosh{\frac{\beta\epsilon_j}{2}\sqrt{1+(\frac{16\xi_j^{\prime
2}y}{\epsilon_j^2})}})}\right).
\end{equation}
The last integral can be evaluated accurately using Laplace's method \cite{Laplace}.  This method makes use of the fact that, if $N \gg 1$, the integral should be dominated by values of $y$ near the maximum of $\phi(y)$, which is determined by the condition
\begin{equation}
\phi^\prime(y) = 0. \label{eq:phip}
\end{equation}
In our case, $\phi^\prime(y)=-\beta\hbar\omega+
\frac{1}{N}\sum_j\frac{4\xi_j^{\prime 2}\beta}{\epsilon_j
\sqrt{1+\frac{16\xi_j^{\prime 2}}{\epsilon_j^2}y}}
\tanh{(\frac{\beta\epsilon_j}{2}\sqrt{1+(\frac{16\xi_j^{\prime
2}y}{\epsilon_j^2})})}$; so eq.\ (\ref{eq:phip}) becomes
\begin{equation}
\frac{\hbar\omega}{4} = \frac{1}{N}\sum_j\frac{\xi_j^{\prime
2}}{\epsilon_j\eta_j}\tanh{\left(\frac{\beta\epsilon_j}{2}\eta_j\right)},
\end{equation}
where $\eta_j=\sqrt{1+\frac{16{\xi_j^{\prime^2}} y}{\epsilon_j^2}}$,
and lies in the range $1 < \eta_j < \infty$. For large $\beta$,
$\tanh{\left(\frac{\beta\epsilon_j}{2}\eta_j\right)} \rightarrow 1$.
Thus, when $\frac{\hbar\omega}{4}>
\frac{1}{N}\sum_j\frac{\xi_j^{\prime 2}}{\epsilon_j}$, $\phi(y)$ is
maximum at $y = 0$. On the other hand, when $\frac{\hbar\omega}{4}<
\frac{1}{N}\sum_j\frac{\xi_j^{\prime 2}}{\epsilon_j}$, the allowed
solutions to eq.\ (\ref{eq:phip}) depend on the value of $\beta$.
When $\beta$ is smaller than a critical value $\beta_c$ given by
\begin{equation}
\frac{\hbar\omega}{4} = \frac{1}{N}\sum_j\frac{\xi_j^{\prime
2}}{\epsilon_j} \tanh{\left(\frac{\beta_c\epsilon_j}{2}\right)},
\label{eq:tc}
\end{equation}
then again $\phi (y)$ is maximum for $y = 0$.  However, if $\beta
>\beta_c$, there is a non-zero solution $y_0$ for $y$ determined by the equation
\begin{equation}
\frac{\hbar\omega}{4} = \frac{1}{N}\sum_j\frac{\xi_j^{\prime
2}}{\epsilon_j\sqrt{1+\frac{16{\xi_j^{\prime^2}}
y_0}{\epsilon_j^2}}}\tanh{\left(\frac{\beta\epsilon_j}{2}\sqrt{1+\frac{16{\xi_j^{\prime^2}}
y_0}{\epsilon_j^2}}\right)} . \label{eq:dDickecon}
\end{equation}

Therefore, we can discuss the statistical mechanics of this model in three different regimes: $(i)$ $\sum_j\frac{\xi_j^{\prime 2}}{N\epsilon_j} \le \frac{\hbar\omega}{4}$; $(ii)$ $\sum_j\frac{\xi_j^{\prime 2}}{N\epsilon_j} \ge
\frac{\hbar\omega}{4}$ and $T\ge T_c$; and $(iii)$ $\sum_j\frac{\xi_j^{\prime 2}}{N\epsilon_j} \ge \frac{\hbar\omega}{4}$ and $T\le T_c$, where $T_c$ is determined by eq.\ (\ref{eq:tc}). In the regimes $(i)$ and $(ii)$, the free energy is given simply by
\begin{equation}
\lim_{N\rightarrow\infty}\frac{F}{N}=-k_BT\phi(y)_{y=0}=-k_BT\frac{1}{N}\sum_j
\ln{(2\cosh{\frac{\beta\epsilon_j}{2}})}
\end{equation}
and the moments of the photon occupation number by
\begin{equation}
\langle \left(\frac{a^{\dagger}a}{N}\right)^k\rangle = (y^k)_{y=0}=
\delta_{k0},
\end{equation}
where the last result is obtained once again by using Laplace's method. On the other hand, in regime $(iii)$,
\begin{equation}
\lim_{N\rightarrow\infty}\frac{F}{N}=
-k_BT\phi(y)_{y=y_0}
\end{equation}
and
\begin{equation}
\langle \left(\frac{a^{\dagger}a}{N}\right)^k\rangle=(y^k)_{y=y_0} \equiv y_0^k
\end{equation}
where $y_0$ is determined by eq.\ (\ref{eq:dDickecon}).

Assuming that all junctions are identical,
$\epsilon_j\rightarrow\epsilon, \xi_j\rightarrow\xi,
\lambda_j\rightarrow\lambda$, the conditions for the critical junction number and the critical temperature become \begin{eqnarray}
N_c^{Dicke}&=& \frac{\hbar\omega\epsilon}{4|\xi |^2} \label{eq:NcDicke},\\
k_BT_c^{Dicke}&=&\frac{1}{\beta_c}=\frac{\epsilon}{2\tanh^{-1}{\left(\frac{\hbar\omega \epsilon}{4N|\xi |^2}\right)}}. \label{eq:TcDicke}
\end{eqnarray} Furthermore, the moments $\langle (a^{\dag}a)^k\rangle=N y_0^k$ are obtained from
\begin{equation}
\frac{\hbar\omega\epsilon}
{4N\xi^2}\sqrt{1+\frac{16N\xi^2y_0}{\epsilon^2}}=\tanh{\frac{\beta\epsilon}{2}
\sqrt{1+\frac{16N\xi^2y_0}
{\epsilon^2}}} .
\label{eq:ConditionDicke}
\end{equation}

\subsection{ \label{sec:MDicke} Statistical mechanics of modified Dicke model using a mean-field approximation}

Next, we consider the modified Dicke model, eq.\ (\ref{eq:MDicke}). The coherence transition can be obtained if we make the following mean-field approximation:
\begin{eqnarray}
a^{\dagger}\sigma_-^j&\sim& \langle a^{\dag}\rangle\sigma_-^j+
\langle \sigma_-^j\rangle a^{\dag}-\langle \sigma_-^j\rangle\langle a^{\dag}\rangle , \\
a\sigma_+^j &\sim& \langle a\rangle\sigma_+^j+
\langle \sigma_+^j\rangle a-\langle \sigma_+^j\rangle\langle a\rangle , \\
\sigma_+^j\sigma_-^k&\sim& \langle\sigma_+^j \rangle\sigma_-^k+
\langle\sigma_-^k \rangle\sigma_+^j-\langle\sigma_+^j \rangle\langle\sigma_-^k\rangle , \\
\sigma_-^j\sigma_+^k&\sim& \langle\sigma_-^j
\rangle\sigma_+^k+\langle\sigma_+^k
\rangle\sigma_-^j-\langle\sigma_-^j
\rangle\langle\sigma_+^k\rangle .
\end{eqnarray}
With the additional assumption that the $\xi_j$'s are real, the Hamiltonian (\ref{eq:MDicke}) separates into a sum of three terms as follows:
\begin{equation}
H^{MDicke}=H_{photon}^{MDicke}+\sum_jH_{Jj}^{MDicke}+H_{c}^{MDicke} ,
\end{equation}
where
\begin{eqnarray}
H_{photon}^{MDicke}&=&\hbar\omega (a^{\dag}a+
\frac{1}{2})+\sum_j\xi_j (\langle\sigma_x^j \rangle
(a+a^{\dag})+i\langle \sigma_y^j\rangle (a-a^{\dag})), \\
\sum_jH_{Jj}^{MDicke}&=& \sum_j\frac{\epsilon_j}{2}\sigma_z^j+
\sum_j\xi_j\left(\langle a^{\dagger}\rangle\sigma_-^j+\langle a\rangle\sigma_+^j\right)+
\sum_j\sum_{k\ne j}\Omega_{jk} (\langle\sigma_+^j \rangle\sigma_-^k+
\langle\sigma_-^k \rangle\sigma_+^j) , \\
H_{c}^{MDicke}&=& -\sum_j\xi_j \left( \langle
a^{\dagger}\rangle\langle\sigma_-^j\rangle+ \langle
a\rangle\langle\sigma_+^j\rangle\right)- \sum_{\langle
jk\rangle}\Omega_{jk} \langle\sigma_+^j \rangle
\langle\sigma_-^k\rangle,
\end{eqnarray}
and we have used the relations
$\langle\sigma_+^j\rangle=\langle\sigma_x^j\rangle+i\langle\sigma_y^j\rangle$ and $\langle\sigma_-^j\rangle=\langle\sigma_x^j\rangle-i\langle\sigma_y^j\rangle$.

The free energy associated with each term in the above Hamiltonian can be evaluated separately. $H^{MDicke}_{photon}$ is the Hamiltonian of a harmonic oscillator displaced in both momentum and position space. Introducing the operators
\begin{equation}
u=\sqrt{\frac{\hbar}{2\omega}}(a+a^{\dag}), \quad
p=i\sqrt{\frac{\hbar\omega}{2}}(a^{\dag}-a),
\end{equation}
we find that $H_{photon}^{MDicke}$ can be rewritten as
\begin{eqnarray}
H_{photon}^{MDicke}&=&\frac{1}{2m} \left(p-p_0 \right)^2+
\frac{m\omega^2}{2}\left(u -u_0
\right)^2 \nonumber \\
& & - \frac{\sum_{jk}\xi_j\xi_k(\langle\sigma_x^j
\rangle\langle\sigma_x^k \rangle+\langle\sigma_y^j
\rangle\langle\sigma_y^k \rangle )}{\hbar\omega}.
\end{eqnarray}
where $p_0 = \sqrt{\frac{2}{\hbar\omega}}
\sum_j\langle\sigma_y^j\rangle\xi_j$ and $u_0 =
-\sqrt{\frac{2}{\hbar\omega^3}}\sum_j
\langle\sigma_x^j\rangle\xi_j$. The corresponding eigenvalues are
\begin{equation}
E_{photon}^{MDicke}(n) = \hbar\omega (n+\frac{1}{2})-
\frac{\sum_{jk}\xi_j\xi_k(\langle\sigma_x^j \rangle
\langle\sigma_x^k \rangle+\langle\sigma_y^j \rangle\langle\sigma_y^k
\rangle )}{\hbar\omega}.
\end{equation}
Also, from the fact that $\langle p \rangle = p_0$ and $\langle u
\rangle = u_0$, we obtain $\langle a\rangle =
-\frac{\sum_j\xi_j\langle\sigma_-^j\rangle}{\hbar\omega}, \quad
\langle a^{\dag}\rangle=-\frac{\sum_j\xi_j\langle\sigma_+^j\rangle}{\hbar\omega}$.

$H^{MDicke}_{Jj}$ is the Hamiltonian of a collection of non-interacting spin-1/2 particles in an applied effective magnetic field (which is not parallel to the $z$ axis). The two eigenvalues of $H^{MDicke}_{Jj}$ are readily found to be
\begin{equation}
E_{Jj}^{MDicke}=\pm\sqrt{\frac{\epsilon_j^2}{4}+4\left(\xi_j\langle
a^{\dag}\rangle+\sum_{k\ne
j}\frac{\Omega_{jk}}{2}\langle\sigma_+^k\rangle\right)\left(\xi_j\langle
a\rangle+\sum_{k\ne
j}\frac{\Omega_{jk}}{2}\langle\sigma_-^k\rangle\right)} .
\end{equation}
Finally, $H^{MDicke}_{c}$ is just a c-number whose expectation value
is just $E_c^{MDicke} = H_c^{MDicke}$.

For $N$ identical junctions, $\epsilon_j\rightarrow \epsilon,
\xi_j\rightarrow\xi$, and $\Omega_{jk}\rightarrow\Omega$. Then, after some algebra, one finds that the ground state energy can be written in terms of a single expectation value $\langle \sigma_\perp\rangle =
[\langle\sigma_x\rangle^2+\langle\sigma_y\rangle^2]^{1/2}$. The result is
\begin{equation}
E_0^{MDicke}(\langle\sigma_{\perp}
\rangle)\!=\!\frac{N^2\xi^2\langle\sigma_{\perp}\rangle^2}{\hbar\omega}
\!\left(\!1\!-\!\frac{(N-1)\hbar\omega\Omega}{2N\xi^2}\right)
-N\sqrt{\frac{\epsilon^2}{4}\!+\!4\langle\sigma_{\perp}
\rangle^2\!\left(\frac{(N-1)}{2}\Omega\!-\!\frac{N\xi^2}{\hbar\omega}\right)^2}.
\end{equation}
$\langle\sigma_\perp\rangle$ is again determined by the requirement that $E_0^{MDicke}(\langle\sigma_{\perp} \rangle)$ be a minimum with respect to $\langle\sigma_{\perp} \rangle$, which leads to \begin{equation} \langle\sigma_{\perp} \rangle^2=1-\frac{1}{4}\left(\frac{\hbar\omega\epsilon}{2N\xi^2-(N-1)\hbar\omega\Omega}\right)^2 .
\label{eq:sigperp}
\end{equation}
The condition $\langle\sigma_{\perp} \rangle^2>0$ leads to the critical number of junctions \begin{equation} N_{c}^{MDicke}=
\frac{\hbar\omega\epsilon}{4\xi^2}
\frac{1-\frac{2\Omega}{\epsilon}}{1-\frac{\Omega\hbar\omega}{2\xi^2}},
\label{eq:NcMDicke}
\end{equation}
above which $\langle\sigma_\perp\rangle^2$ is non-negative. When
$\Omega=0$, this critical number exactly corresponds to the critical number obtained in Section \ref{sec:jose_dicke}. Thus, for the Dicke model, this
MFT yields the same critical junction number as obtained from the coherent state analysis.

At finite $T$, the properties of the modified Dicke model are obtained from the Helmholtz free energy $F^{MDicke}$. An analysis similar to that at $T = 0$ again allows $F^{MDicke}$ to be written as the sum of three terms, which for $N$ identical junctions may be written
\begin{equation}
F^{MDicke}=-k_BT\ln Z_{photon}^{MDicke} -
Nk_BT\ln(Z_J^{MDicke})+E_{constant}^{MDicke},
\end{equation}
where $Z_{photon}^{MDicke} =
\frac{1}{2\sinh{\frac{\beta\hbar\omega}{2}}}$; $Z_{J}^{MDicke} =
2\cosh{\left(\beta\sqrt{\frac{\epsilon^2}{4}+4
\langle\sigma_{\perp}\rangle^2\left(\frac{(N-1)}{2}\Omega-
\frac{N\xi^2}{\hbar\omega}\right)^2}\right)}$, and $E_{constant}^{MDicke}
= \frac{N^2\xi^2\langle\sigma_{\perp}\rangle^2}
{\hbar\omega}\left(1-\frac{(N-1)\Omega\hbar\omega}{2N\xi^2}\right)$, which also includes constant contributions from photon and junction terms.

As at $T = 0$, the optimal value of $\langle\sigma_{\perp} \rangle$ at finite $T$ is obtained by minimizing $F^{MDicke}$ with respect to $\langle\sigma_\perp\rangle$, which leads to the following relation for $\langle\sigma_\perp\rangle$:
\begin{equation}
\frac{\hbar\omega\epsilon\eta}{4N\xi^2\left(1-\frac{(N-1)\hbar\omega\Omega}{2N\xi^2}\right)}
=\tanh\left[\frac{\beta\epsilon}{2}\eta\right],
\label{eq:ConditionMDicke}
\end{equation}
where $\eta=\sqrt{1+\frac{16N^2\xi^4\langle\sigma_{\perp}\rangle^2} {\epsilon^2(\hbar\omega)^2}\left(1-\frac{(N-1)\hbar\omega\Omega}{2N\xi^2}\right)^2}$.
The critical temperature $T_c^{MDicke}$ for this modified Dicke model is again determined by the requirement that $\langle\sigma_\perp\rangle^2 > 0$, and is given by
\begin{equation}
k_BT_c^{MDicke}=\frac{\epsilon}{2\tanh^{-1}
{\left(\frac{\hbar\omega\epsilon}{4N\xi^2}\frac{1}{\left(1-\frac{\hbar\omega (N-1)\Omega}
{2N\xi^2}\right)}\right)}}. \label{eq:TcMDicke}
\end{equation}
$k_BT_c^{MDicke}=0$ for
\begin{equation}
\Omega >\Omega_{max} = \frac{2N\xi^2}{(N-1)\hbar\omega}\left(1-\frac{\hbar\omega}{4N\xi^2}\right).
\end{equation}
When $\Omega=0$, eq.\ (\ref{eq:TcMDicke}) for $T_c$ reduces to eq.\
(\ref{eq:TcDicke}), provided we assume $\langle
a^{\dag}\rangle\langle a\rangle = \langle a^{\dag}a\rangle$ and use the relation
$\langle\sigma_{\perp}\rangle^2=\left(\frac{\hbar\omega}{N\xi}\right)^2\langle
a^{\dag}\rangle\langle a\rangle$. Thus, both MFT and the coherent state expansion lead to the same thermodynamic properties for the Dicke model.

\section{Numerical Results \label{sec:jose_result}}
We have carried out several illustrative numerical calculations using the models and approximations of Sections \ref{sec:MFT} and \ref{sec:jose_dicke}. For the MFT of Section \ref{sec:MFT}, these results are obtained by minimizing the mean-field Helmholtz free energy, eq.\ (\ref{eq:Fm}), with respect to $\lambda$ at each $T$ for fixed $g$ and $\bar{n}$.
In all our calculations, we have taken $\hbar\omega=0.15U, J=0.2U, g=0.1$, and $N=110$; other parameters are described below. It is convenient to introduce a dimensionless temperature $t\equiv \frac{k_BT}{U}$. Except for $g$, these are in the same ratios as in recent eperiments of Ref.\cite{Schuster2005} using the correspondence, $(\hbar\omega,U/8,J)$ in our notation to $(\hbar\omega_r= 6.0\mathrm{GHz}$, $E_C\sim 5.0\mathrm{GHz}$, and $E_{J,max}\sim 8.0\mathrm{GHz})$ in the experiment. However, we have used a much larger value of $g$, in order to see the transition to a coherent state at a reasonable value of $N$.

\begin{figure}
\begin{center}
\includegraphics[height=8cm,width=10cm]{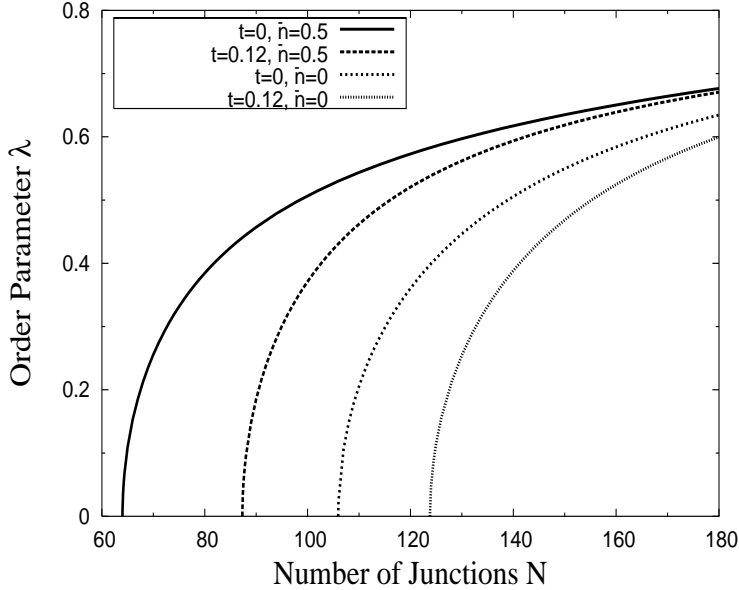}
\caption{\label{fig:rvsn} Coherence order parameter $\lambda(N,t)$, plotted as a function of the number of junctions $N$, for $\bar{n}=0$ and $0.5$ at values of the scaled temperature $t=0$ and $0.12$, as indicated in the legend, using $J=0.2U$, $\hbar\omega=0.15U$, and $g=0.1$.}
\end{center}
\end{figure}

\begin{figure}
\begin{center}
\includegraphics[height=8cm,width=10cm]{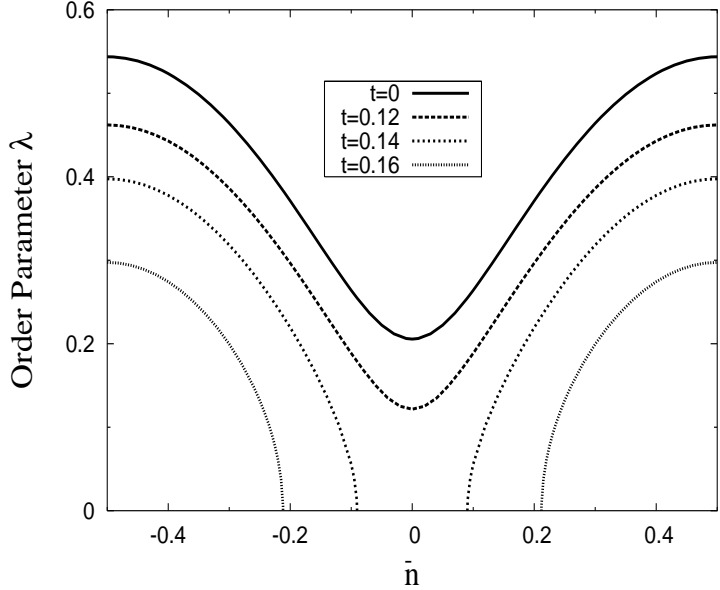}
\caption{\label{fig:rvsnb} Coherence order parameter
$\lambda(\bar{n}, t)$, at several values of the scaled temperature
$t=0, 0.12, 0.14$ and $0.16$, as indicated in the legend, using
$J=0.2U$, $\hbar\omega=0.15U, g=0.1$, and $N=110$.}
\end{center}
\end{figure}

\begin{figure}
\begin{center}
\includegraphics[height=8cm,width=10cm]{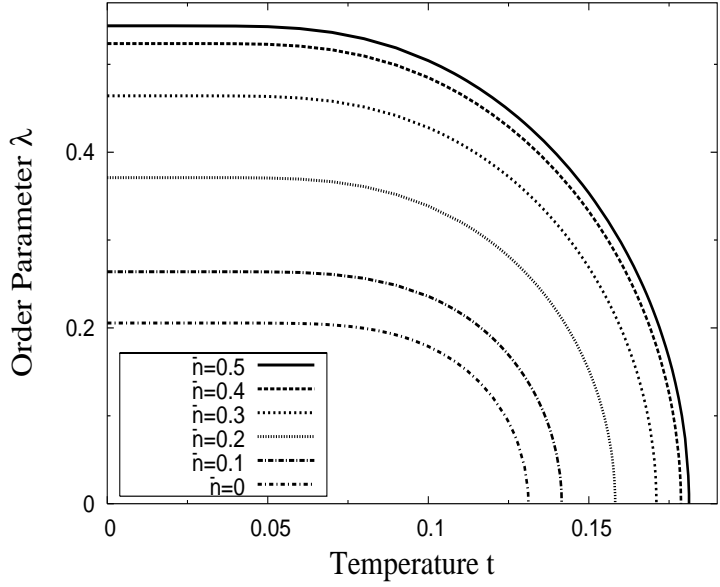}
\caption{\label{fig:rvst} Temperature dependence of coherence order parameter $\lambda(t)$ for $\bar{n} = 0, 0.1, 0.2, 0.3, 0.4$ and $0.5$, using $J=0.2U$, $\hbar\omega=0.15U, g=0.1$, and $N=110$.}
\end{center}
\end{figure}

\begin{figure}
\begin{center}
\includegraphics[height=8cm,width=10cm]{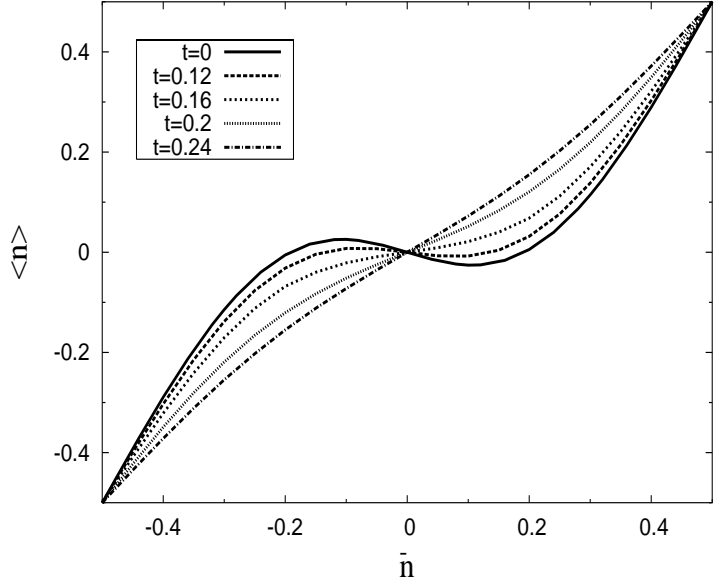}
\caption{\label{fig:anvsnb} Average Cooper pair number difference
$\langle n \rangle$ at $t=k_BT/U=0, 0.12, 0.16, 0.2$, and $0.24$. In all cases, $J=0.2U$, $\hbar\omega=0.15U, g=0.1$, and $N=110$ where $t_c(\bar{n}=0)=0.124$ and $t_c(\bar{n}=0.5)=0.181$.}
\end{center}
\end{figure}

\begin{figure}
\begin{center}
\includegraphics[height=7.5cm,width=10cm]{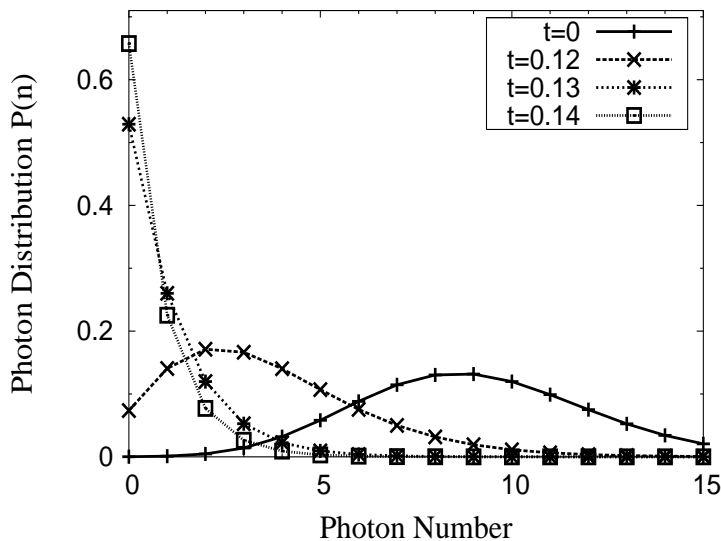}
\caption{\label{fig:pvsn} Photon number distribution $P(n)$ for the Josephson junction model at various temperatures: $t=0, 0.12, 0.13$, and $0.14$. In all cases, $J=0.2U$, $\hbar\omega=0.15U, g=0.1$, $N=110$, and $\bar{n}=0$. For these parameters, $t_c=0.131$. $P(n)$ represents the probability that there are exactly $n$ photons in the cavity mode.}
\end{center}
\end{figure}

Figs.~\ref{fig:rvsn}, \ref{fig:rvsnb}, \ref{fig:rvst}, \ref{fig:anvsnb}, and \ref{fig:pvsn} show mean-field results for the Hamiltonian of Section \ref{sec:jose_model}, including all Josephson levels. First, we consider the coherence order parameter $\lambda(N,T)$ assuming $g$ independent of $N$. Fig.~\ref{fig:rvsn} shows $\lambda(N,T)$ for $\bar{n} = 0$ and $0.5$ at $t=0$ and $0.12$.
In all cases, there is obviously a threshold number of junctions $N_c(t)$ below which $\lambda$ vanishes, and above which $\lambda\neq 0$. For a sufficiently large $N$, $\lambda \rightarrow 1$, signaling complete phase locking. Fig.~\ref{fig:rvsn} shows that both $\lambda(t)$ and $N_c(t)$ decrease with increasing $t$, and that $\bar{n} = 0.5$ leads to a larger $\lambda$ at fixed $t$ than does $\bar{n} = 0$. Both features are intuitively reasonable, since at $\bar{n} = 0.5$, the two lowest states of the junction have only a small gap, making it easier to couple the junction to the cavity.

Fig.~\ref{fig:rvsnb} shows $\lambda(\bar{n}, t)$ as a function of $\bar{n}$, which is related to the voltage across the Josephson array, at $t=0, 0.12, 0.14$ and $0.16$.
Since $\lambda(\bar{n}, t)$ is a periodic function of $\bar{n}$ with period unity, we plot only the range $-0.5\le\bar{n}\le 0.5$. All the plots of Fig.~\ref{fig:rvsnb} show that, for any choice of the other parameters, $\lambda$ is maximum at $\bar{n}=0.5$. The plots also show that there exist values of $N$ such that the array is coherent for some non-zero values of
$\bar{n}$ even if it is incoherent at $\bar{n} = 0$. Finally, Fig.~\ref{fig:rvsnb} shows, as expected and as is also shown in Fig.~\ref{fig:rvsn}, that the effect of increasing $t$ at fixed $N$ and $g$ is to suppress $\lambda$.

The temperature dependence of $\lambda(\bar{n},t)$ is plotted versus $t$ in Fig.~\ref{fig:rvst} for different values of $\bar{n}$.
In all cases, there is a critical temperature $t_c$ above which $\lambda=0$. Note also that, as $t \rightarrow t_c$ from below, $\lambda \rightarrow 0$ {\em continuously}. This behavior is a hallmark of a continuous phase transition. However, since there are only a finite number of junctions, the transition is not a true thermodynamic phase transition.

We have also calculated the average Cooper pair difference $\langle n_j\rangle$ across the j$^{th}$ junction, within the mean-field approximation. Since all the junctions are assumed identical, $\langle n_j\rangle$ is independent of $j$ and may be denoted $\langle n \rangle$. $\langle n\rangle$ is related to the voltage drop $V$ across a junction by $CV/2e =\langle n\rangle$, and can be calculated from the relation
\begin{equation}
\langle n\rangle=\frac{\sum_{k=0,\pm 1\pm 2...}e^{-\beta
E(\nu=2\bar{n}+2k,q)}\langle n\rangle_k}{\sum_{k=0, \pm 1, \pm
2,...}e^{-\beta E(\nu=2\bar{n}+2k,q)}}.
\end{equation}
where $\langle n\rangle_k$, the expectation value of the operator $n$ in state $k$, is
\begin{equation}
\langle n\rangle_k=\int_0^{2\pi}\psi^*_{\nu=2\bar{n}+2k}(\phi)
 \left(i\frac{\partial}{\partial\phi}\right)\psi_{\nu=2\bar{n}+2k}
 (\phi)d\phi = \bar{n}+i\int_0^\pi
 y_{\nu=2\bar{n}+2k}(v)\frac{dy_{\nu=2\bar{n}+2k}(v)}{dv}dv.
\end{equation}
The last expression is obtained using the relations $n=i\frac{\partial}{\partial\phi}=
\frac{i}{2}\frac{\partial}{\partial v}$ and $\psi_\nu(\phi)=
e^{-i\bar{n}(\phi-\alpha)}u_\nu(\phi-\alpha)=e^{-2i\bar{n}v}y_\nu(v)$.
In Fig.~\ref{fig:anvsnb}, we show this calculated $\langle n\rangle(\bar{n}, t)$ versus $\bar{n}$ for several values of $t$. Note that, $t\gg t_c$, $\langle n\rangle\sim\bar{n}$.

Next, we discuss the temperature dependence of the photon probability distribution $P(n,t)$ calculated in this mean-field approximation. $P(n,t)$ is defined simply as the probability that the cavity contains exactly $n$ photons at temperature $t$. Since there is no coherence for $t>t_c$, $P(n)$ is given simply by the usual Bose distribution with zero chemical potential:
\begin{equation} P(n)=\frac{e^{-\beta\hbar\omega(n+1/2)}}{\sum_{l=0}^{\infty}{e^{-\beta\hbar\omega(l+1/2)}}}
=\frac{e^{-\beta\hbar\omega(n+1/2)}}{2\sinh{\beta\hbar\omega/2}}
\end{equation}
[eq.\ (\ref{eq:hphotmf})]
\begin{equation}
P(n) = \langle n|\rho |n\rangle.
\end{equation}
Using the solutions of $H_{photon}^m$,
\begin{equation}
\rho = \frac{1}{Z_{photon}^m}\sum_{l=0}^{\infty}e^{-\beta E_l}e^{-ip\langle
x\rangle/\hbar}|l\rangle\langle l|e^{ip\langle x\rangle/\hbar} ,
\label{eq:rho}
\end{equation}
where $Z_{photon}^m$ is the partition function corresponding to $H_{photon}^m$.
Thus, we obtain,
\begin{equation}
P(n)Z_{photon}^m=\sum_{l=0}^{\infty}e^{-\beta E_l}
\langle n|e^{-ip\langle x\rangle /\hbar}|l\rangle
\langle l|e^{ip\langle x\rangle /\hbar}|n\rangle =\sum_{l=0}^{\infty}|
\langle n |e^{-ip\langle x\rangle /\hbar}|l\rangle|^2 e^{-\beta E_l} ,
\end{equation}
where
\begin{equation}
\langle n |e^{-ip\langle x\rangle /\hbar}|l\rangle
=\langle n|\left[\int_{-\infty}^{\infty}|x\rangle\langle x|\right]|e^{-ip\langle x\rangle /\hbar}|l\rangle
= \int_{-\infty}^{\infty}\psi_n^*(x+\langle x\rangle)\psi_l(x) dx .
\end{equation}
For the consideration of only $l=0$ term, the probability function $P(n)$ corresponds to
\begin{equation}
P_0(n)=\frac{|\alpha|^{2n}}{n!}e^{-|\alpha|^2} ,
\end{equation}
where $P_0(n)$ has a maximum at $n=|\alpha|^2$.

Fig.~\ref{fig:pvsn} shows this photon distribution at $t\equiv k_BT/U=0, 0.12, 0.13$, and $0.14$ where $t_c=0.131$. From low $t$ up
to near $t_c$, $P(n,t)$ is substantial over a wide range of $n$, but
for $t\ge t_c$, the population of the photon state with $n = 0$ rapidly increases.

\begin{figure}
\begin{center}
\includegraphics[height=8cm,width=10cm]{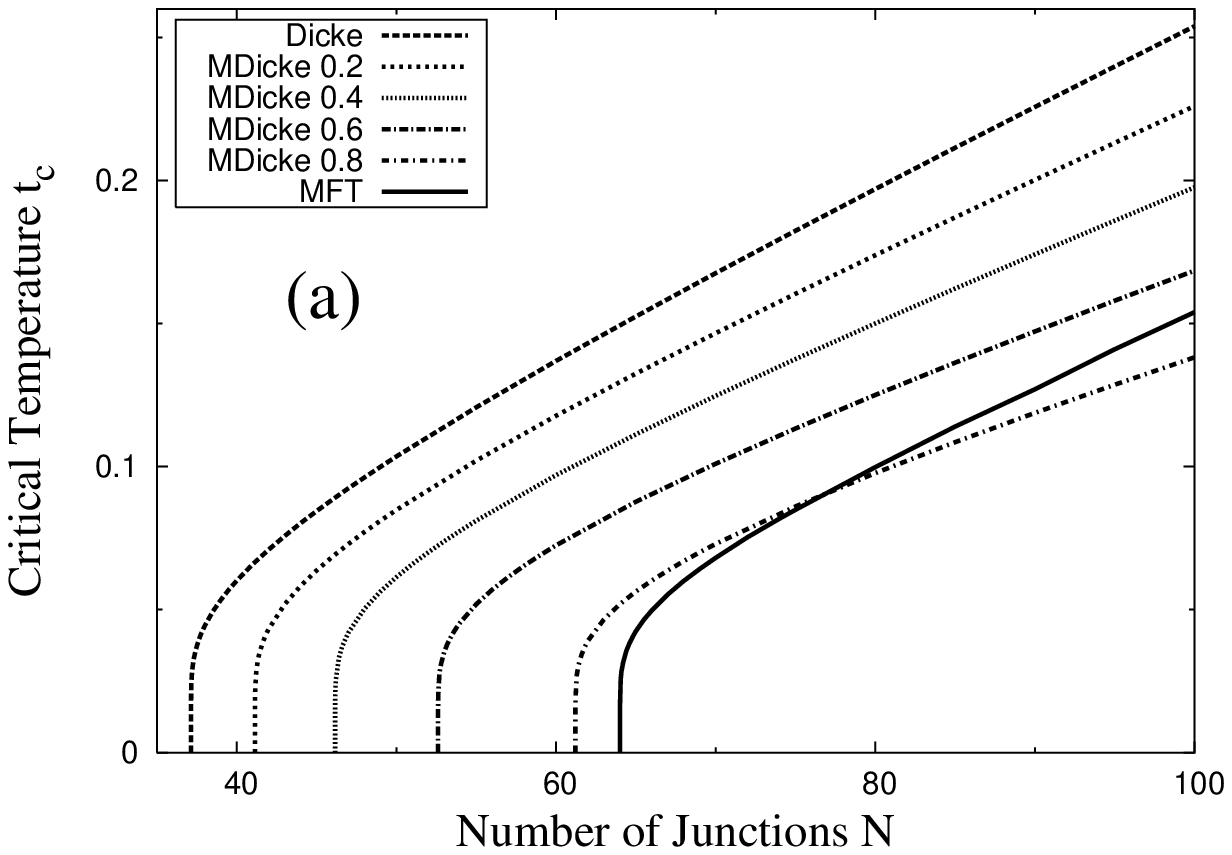}
\includegraphics[height=8cm,width=10cm]{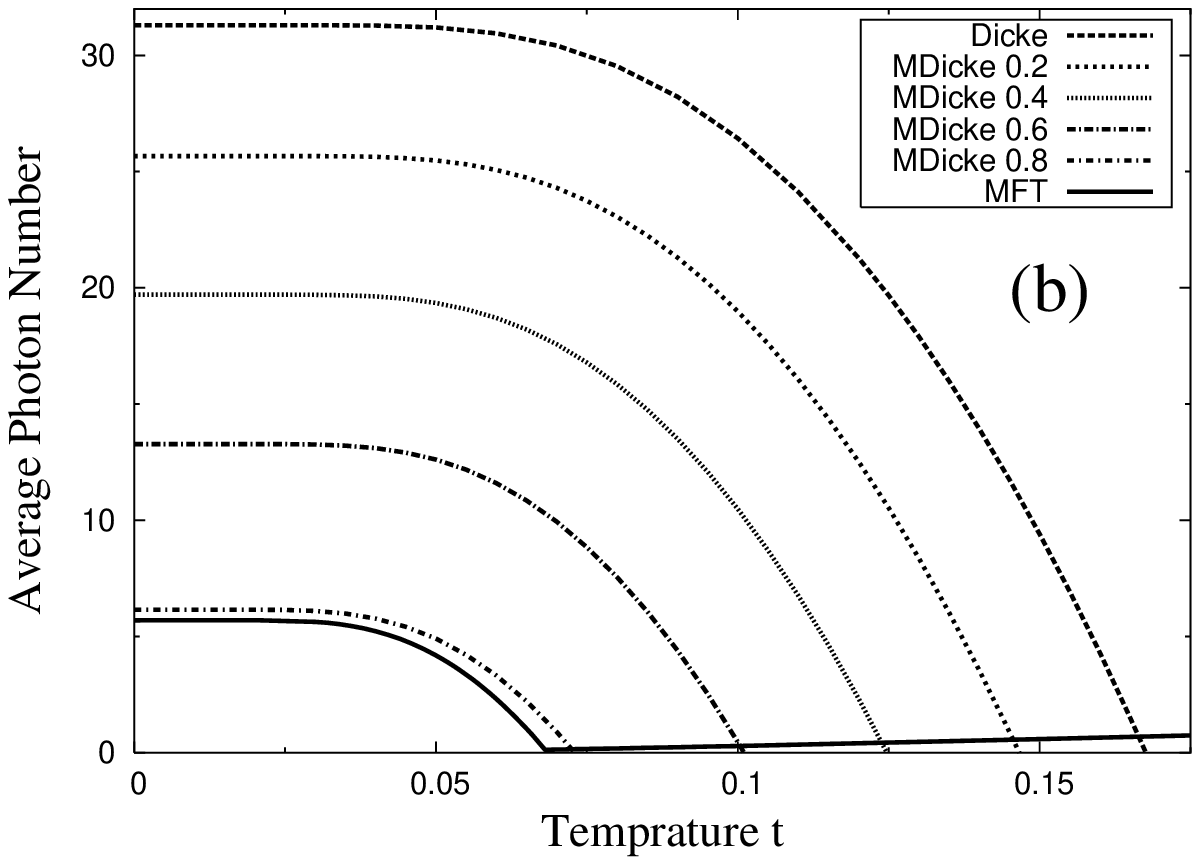}
\caption{\label{fig:comp} Comparison between the predictions of the mean-field approximation, the Dicke model, and the modified Dicke model for the critical temperature [part (a)] and the average photon number, $\langle a^\dagger a\rangle(t)$ for $N=70$ [part (b)], at $\bar{n}=0.5$.  For the Josephson junction model, we use the parameters $J=0.2U$, $\hbar\omega=0.15U$, and $g=0.1$; these lead to $N_c^m( 0)=64.0$ and $t_c^m=0.0681$. The corresponding critical numbers for the Dicke model are $N_c^{Dicke}(0)=37.1$ and
$t_c^{Dicke}=0.168$. For the modified Dicke model, we show plots with $\Omega\hbar\omega/|\xi |^2 = 0.2$, $0.4$, $0.6$ and $0.8$.}
\end{center}
\end{figure}

Next, we compare the mean-field results of Section \ref{sec:jose_model} (which includes all junction levels) to the results of Section \ref{sec:jose_dicke} for the Dicke model and the modified Dicke model. We consider specifically $\bar{n}=0.5$; at this value of $\bar{n}$, the two-level approximation may be best, because the two lowest junction levels are separated by the largest gap from the higher levels. In order to compare the three models, we plot $t_c(N)$ in Fig.~\ref{fig:comp}(a), and the average photon number $\langle a^\dag a\rangle(t)$ in Fig.~\ref{fig:comp}(b). For comparison purposes, we choose the Dicke parameter $\xi = -\frac{gJ}{\sqrt{2}}$, and the Dicke parameter $\epsilon =E(2-2\bar{n},q)-E(-2\bar{n},q)=(U/8)[a(2-2\bar{n},q)-a(-2\bar{n},q)]$
\cite{Comment1}. Also, we treat $\Omega$ simply as a parameter determined by best fitting to the results of the MFT.  We denote the critical number of junctions and the critical temperature of the MFT by $N_c^m(t)$, $t_c^m$; of the Dicke model, by $N_c^{Dicke}(t)$, $t_c^{Dicke}$; and of the modified Dicke model, by $N_c^{MDicke}(t)$, $t_c^{MDicke}$. The $t_c^m$'s are obtained numerically; the other $t_c$'s are obtained from eq.\ (\ref{eq:TcDicke}) for $t_c^{Dicke}$ and from eq.\ (\ref{eq:TcMDicke}) for $t_c^{MDicke}$. In the mean-field case, when the coherence order parameter vanishes we just have the
Bose result for the average photon occupation number:
\begin{equation}
\langle a^{\dag}a\rangle_m(t>t_c^m) =
\frac{1}{e^{\frac{\hbar\omega}{k_BT}}-1}.
\end{equation}
On the other hand, in the coherent state, we have
\begin{equation}
\langle a^{\dag}a\rangle_m(t<t_c^m)=\sum_{n=0}^{\infty} n P(n) .
\end{equation}
For the other two models, $\langle a^{\dag}a\rangle_{Dicke}$ and $\langle a^{\dag}a\rangle_{MDicke}$ are calculated from the conditions (\ref{eq:ConditionDicke}) and (\ref{eq:ConditionMDicke}) with $\langle\sigma_{\perp}\rangle^2=(\frac{\hbar\omega}{N\xi})^2\langle a^{\dag} a\rangle$, respectively.

In Fig.\ \ref{fig:comp}, we plot $t_c(N)$ and the average photon number $\langle a^\dag a\rangle (t)$, for the three models.
To compare the MFT with the modified Dicke model, we have considered four choices for $\Omega$, corresponding to $\Omega\hbar\omega/|\xi |^2 =0.2$, $0.4$, $0.6$ and $0.8$. All three models show the same qualitative behavior, i.\ e., a transition from coherence to incoherence with decreasing $N$ or increasing $t$. The solutions of the Dicke and modified Dicke models are qualitatively in good agreement with that of the MFT; however, the solution of the modified Dicke model with $\Omega=0.8|\xi |^2/(\hbar\omega)$ agrees better with MFT than do any of the other three.

The behavior of $\langle a^\dag a\rangle(t)$ differs somewhat among the three models. For the MFT model, $\langle a^{\dag}a\rangle(t) \rightarrow 0$ as $t \rightarrow t_c^m$ from below, and remains very small, but non-zero, for $t > t_c^m$. On the other hand, in both the Dicke and modified Dicke models, $\langle a^\dag a\rangle$ reaches exactly zero at $t = t_c$ and remains zero for $t
> t_c$. The most conspicuous qualitative difference between the two models occurs at large $t$, where the Dicke and modified Dicke models give $\langle a^\dag a\rangle(t) = 0$, while $\langle a^\dag a \rangle(t)$ in the MFT increases with increasing $t$ according to the Bose distribution. This discrepancy probably occurs because the first two models include only two levels per junction, while the MFT of Section \ref{sec:jose_model} treats a many-level system.

\section{Thermodynamic Limit \label{sec:jose_limit}}
We now make more precise the connection between the Josephson-cavity model and the Dicke and modified Dicke models in the thermodynamic limit. We first consider the Dicke model at $T = 0$. For $N$ identical two-level systems, the condition for the onset of coherence at $T= 0$ is given by eq.\ (\ref{eq:NcDicke}). With the assumption $\xi = \tilde{\xi}/\sqrt{N}$, this condition becomes
\begin{equation}
\frac{4\tilde{\xi}^2}{\epsilon} = \hbar\omega. \label{eq:cohdicke}
\end{equation}

To map the Josephson-cavity model onto the Dicke model, we assume that the Josephson coupling parameter $g \propto 1/\sqrt{N}$. As mentioned earlier, this assumption seems reasonable in the thermodynamic limit, since according to eq.\ (\ref{eq:gj}), $g \propto 1/\sqrt{V}$ for fixed cavity shape. Writing $g =
\tilde{g}/\sqrt{N}$, we may express the Josephson coherence condition (\ref{eq:nc0e}) as \begin{equation} |a^\prime (0)|\tilde{g}^2J = \hbar\omega \label{eq:cohjos}.
\end{equation}

We now show that this condition reduces to eq.\ ({\ref{eq:cohdicke}) in the limit $J \ll U$. The eigenvalue derivative $|a^\prime (0)|$ can be obtained approximately in this limit by differentiating the right-hand side of eq.\ (\ref{eq:analyt}) with respect to $q$. Substituting back into eq.\ (\ref{eq:cohjos}), we obtain
\begin{equation}
\frac{4(\tilde{g} J)^2}{\sqrt{U^2(1 - 4\bar{n}^2)^2 + 4J^2}} =
\hbar\omega. \label{eq:60}
\end{equation}
We can also compute the splitting between the ground and first excited states. If $0 < \bar{n} < 1/2$, and $J \ll U$, it is easily shown that the splitting $\Delta E$ between the ground and first excited states of the Josephson junction is equal $\Delta E \sim \frac{1}{2}\left[U^2(1 - 2\bar{n})^2 + 4J^2\right]^{1/2}$\cite{Makhlin2001}. Using the approximation, $1-2\bar{n}\sim 1-4\bar{n}^2$, we can rewrite the coherence condition (\ref{eq:60}) as \begin{equation} \frac{2\tilde{g}^2J^2}{\Delta E} = \hbar\omega. \label{eq:cohjos1}
\end{equation}
This condition is identical to eq.\ (\ref{eq:cohdicke}), with the identification $\epsilon \leftrightarrow \Delta E$, $\tilde{\xi} \leftrightarrow -\frac{\tilde{g} J}{\sqrt{2}}$.
The parameter $\Omega$ has primarily a quantitative effect on the coherence transition in this model. As discussed earlier, in order for the modified Dicke model to be well-behaved in the thermodynamic limit, $\Omega$ must vary as $1/N$. Therefore, we write $\Omega = 2\tilde{\Omega}/(N-1)$; we use $N-1$ rather than $N$ since each two-level system interacts with $N-1$ others. Substituting this relation into eq.\ (\ref{eq:sigperp}), and again using $\xi = \tilde{\xi}/\sqrt{N}$, we find that eq.\ (\ref{eq:cohdicke}) is replaced by
\begin{equation}
\frac{4(\tilde{\xi}^2 - \hbar\omega\tilde{\Omega})}{\epsilon} =
\hbar\omega. \label{eq:mcohdicke}
\end{equation}
If the left-hand side is larger than $\hbar\omega$, the system is coherent at $T = 0$ in the thermodynamic limit; otherwise, it is not. Thus, a positive $\tilde{\Omega}$ actually inhibits coherence at $T = 0$ (not unexpectedly, since a positive $\tilde{\Omega}$ represents a repulsive interaction).

Similarly, one can recalculate $k_BT_c^{MDicke}$ [eq.\ (\ref{eq:TcMDicke}) using the above $N$-dependence of $\xi$ and $\Omega$, with the result
\begin{equation} k_BT^{MDicke}_c =
\frac{\epsilon}{2\tanh^{-1}\left(\frac{\hbar\omega\epsilon}{4\tilde{\xi}^2}\frac{1}{1
- \frac{\hbar\omega\tilde{\Omega}}{\tilde{\xi}^2}}\right)}.
\label{eq:TcMDicke1}
\end{equation}
Thus, for given values of $\omega$, $\epsilon$, and $\tilde{\xi}$, the coherence transition temperature is {\em reduced} by a finite $\tilde{\Omega}$, showing that this form of direct interaction between junctions inhibits coherence in the modified Dicke model.

\section{Discussion \label{sec:jose_discussion}}

In this paper, we have calculated the equilibrium properties of an array of identical Josephson junctions coupled to a single-mode electromagnetic cavity at temperature $T$, by generalizing a $T=0$ MFT \cite{Harbaugh2000}. Within the MFT, this system shows a continuous transition between coherence and incoherence at a critical temperature $T_c$, provided that the number of junctions $N>N_c$. We have also compared our mean-field results to the solutions of the Dicke and modified Dicke models. When the parameters of the Dicke model are adjusted to match those of the Josephson-cavity system, the two approaches agree qualitatively.

Next, we briefly discuss the expected accuracy of our mean-field approach, used in Sections \ref{sec:jose_model} and \ref{sec:jose_dicke}. The MFT appears reasonable, because for both models, all the junctions (or all the two-level systems) interact with the {\em same} harmonic mode and hence, in effect, with all the other junctions or two-level systems. Since each junction or two-level system effectively has many ``neighbors'', there are only small fluctuations in the environment of each about its mean, provided that $N$ is sufficiently large. Thus, the mean-field approach should work well at large $N$. In support of this picture, we have shown that, when the mean-field approach is applied to the Dicke model, it produces the exact result (obtained from a coherent state expansion).

Besides the mean-field approximation in Section \ref{sec:jose_model}, we have expanded the Josephson coupling in powers of the interaction parameter $g(a +
a^\dag)$. The value of this quantity can be estimated as follows.
$g(a + a^\dag)\sim g\langle a+a^\dag\rangle=2Jg^2N\lambda/(\hbar\omega) =
2J\tilde{g}^2\lambda/(\hbar\omega)$ if we assume $g =
\tilde{g}/\sqrt{N}$. Since $\lambda \ll 1$ is small near $T_c$, this approximation is accurate in this regime, but may break down deep in the coherent regime. Therefore, a more accurate approach than that used in Section \ref{sec:jose_model}, may be desirable in order to treat the entire regime $0 < t < t_c$.

We briefly comment on the nature of the coherence transition emerging from our mean-field approach. This approach produces a {\em continuous} transition, i.e., the coherence order parameter $\lambda$ varies continuously with $t$. By contrast, another recent calculation \cite{Lee2004} finds a {\em first-order} transition, in which there is a discontinuous jump in the order parameter at the
superradiant transition. Their Hamiltonian also has the form of a generalized Dicke model, but slightly different from ours:
\begin{equation}
H_{lee} = a^{\dag}a +
\sum_{j=1}^N\left\{\frac{\lambda}{2\sqrt{N}}(a+a^\dag)(\sigma_+^j+\sigma_-^j)
+ \frac{\epsilon}{2}\sigma_z^j-J\sigma_y^j\sigma_y^{j+1}\right\},
\end{equation}
where $\sigma_+^j$, $\sigma_i^j$, and $\sigma_z^j$ are the usual Pauli spin operators for the j$^{th}$ two-level system. Our mean-field treatment of our own generalized Dicke model does not give a first-order transition. We speculate that the difference is due to a real distinction between the two models: $H_{lee}$ has only nearest-neighbor interactions between spins, in addition to the usual Dicke-type model, whereas our modified Dicke Hamiltonian has an additional term which is {\em long-range}. Perhaps the long-range nature of this additional term helps to maintain the continuous nature of the coherence transition, as well as the accuracy of MFT.

Finally, we discuss how our model could be generalized in order to make it more a more realistic basis for treating Josephson arrays in a cavity. For real systems, there are other factors affecting Josephson junctions besides those included here. For example, there are effects due to dissipation, either due to the finite Q of the cavity, or a finite dissipation within individual junctions. Both effects can be treated by considering the Josephson junctions as coupled to appropriate baths of harmonic oscillators \cite{Leggett1987, Weiss1999}. When these dissipative degrees of freedom are properly included, the nature of the coherence transition may be changed. We plan to include some of these effects, as well as the effects of disorder, in a future publication.

\section{Acknowledgments.} This work was supported by the National
Science Foundation through Grant DMR04-13395.


\begin{thebibliography}{99}
\bibitem{Comment1}
From the discussion in Section V, when only the two lowest junction
energy levels are retained, $\epsilon \leftrightarrow \Delta E$,
$\tilde{\xi} \leftrightarrow -\frac{\tilde{g} J}{\sqrt{2}}$. Thus, $(U/8)[a(2-2\bar{n}, q)-a(-2\bar{n}, q)]$ is the splitting
between the lowest two levels of the full Josephson-cavity model and
$\xi=-\frac{gJ}{\sqrt{2}}$, within the mean-field approximation.
\end{thebibliography}
\end{document}